\documentclass[]{spie}  
\usepackage[colorlinks,linkcolor=blue]{hyperref}
\usepackage{setspace}
 
\usepackage{amsmath,amsfonts,amssymb}
\usepackage{graphicx}
\usepackage{multirow}

\usepackage{booktabs}

\usepackage{hyperref}

\hypersetup{hidelinks,
	colorlinks=true,
	allcolors=blue,
	pdfstartview=Fit,
	breaklinks=true}

\title{AmbientCycleGAN for Establishing Interpretable Stochastic Object Models Based on Mathematical Phantoms and Medical Imaging Measurements}

\author[a]{Xichen Xu}
\author[b]{Wentao Chen}
\author[a,b]{Weimin Zhou}

\affil[a]{Global Institute of Future Technology, Shanghai Jiao Tong University, Shanghai, China}
\affil[b]{University of Michigan-Shanghai Jiao Tong University Joint Institute, Shanghai Jiao Tong University, Shanghai, China}

\authorinfo{Further author information: (Send correspondence to Weimin Zhou.)\\Weimin Zhou: E-mail: weimin.zhou@sjtu.edu.cn}

\begin{document} 
	\maketitle
	
\begin{abstract}
Medical imaging systems that are designed for producing diagnostically informative images should be objectively assessed via task-based measures of image quality (IQ). Ideally, computation of task-based measures of IQ needs to account for all sources of randomness in the measurement data, including the variability in the ensemble of objects to be imaged. To address this need, stochastic object models (SOMs) that can generate an ensemble of synthesized objects or phantoms can be employed. Various mathematical SOMs or phantoms were developed that can interpretably synthesize objects, such as lumpy object models and parameterized torso phantoms. However, such SOMs that are purely mathematically defined may not be able to comprehensively capture realistic object variations. To establish realistic SOMs, it is desirable to use experimental data. An augmented generative adversarial network (GAN), AmbientGAN, was recently proposed for establishing SOMs from medical imaging measurements. However, it remains unclear to which extent the AmbientGAN-produced objects can be interpretably controlled. This work introduces a novel approach called AmbientCycleGAN that translates mathematical SOMs to realistic SOMs by use of noisy measurement data. Numerical studies that consider clustered lumpy background (CLB) models and real mammograms are conducted. It is demonstrated that our proposed method can stably establish SOMs based on mathematical models and noisy measurement data. Moreover, the ability of the proposed AmbientCycleGAN to interpretably control image features in the synthesized objects is investigated.
\end{abstract}
	
	
\section{INTRODUCTION}
\label{sec:intro}  
	
When evaluating medical imaging systems, all sources of randomness in measurement data should be accounted for. One important randomness is the variability in the ensemble of objects to be imaged \cite{barrett2013foundations}. Stochastic object model (SOM) that can generate an ensemble of synthesized objects represents an useful tool to provide an \textit{in silico} representation of the object variability, which can be employed for computing task-based measures of image quality\cite{zhou2019approximating, zhou2020approximating,li2021hybrid ,zhou2023Ideal,granstedt2023approximating}.
There has been a long-standing tradition of exploring ways to establish SOMs for synthesizing objects or phantoms. Various mathematical SOMs have been proposed, such as lumpy object models \cite{rolland1992effect}, clustered lumpy models \cite{clb}, binary texture models \cite{abbey2008ideal}, and parameterized torso phantoms \cite{he2008toward}. However, to capture realistic anatomical variations and textures, it is desirable to establish SOMs by use of experimental data. Recently, Zhou \emph{et al.} investigated the use of an augmented generative adversarial network (GAN) architecture, AmbientGAN, to establish realistic SOMs from noisy measurement data \cite{zhouAmbJMI}. However, learning a realistic SOM that possesses the ability to interpretably control synthesized objects remains a challenge.
	
An important deep generative model named Cycle-Consistent Adversarial networks (CycleGAN) \cite{CycleGAN} was developed to learn mappings that can translate between different image styles, and has been successfully employed to perform many unpaired image-to-image translation tasks. By training a CycleGAN that translates between mathematical phantoms and a set of real objects, one may interpretably control features of the synthesized objects by manipulating the parameters of the mathematical phantoms.
Together, the input mathematical phantoms and the trained CycleGAN may represent an interpretable SOM for generating realistic objects. 
However, because medical imaging systems acquire measurement data that are contaminated by noise, CycleGANs can not be directly trained for this purpose.

This study introduces a novel method, AmbientCycleGAN, that can be trained on noisy measurement data for translating phantoms produced by pre-existing mathematical SOMs to realistic objects. Preliminary studies that considered clustered lumpy background (CLB) and real mammograms were conducted. It has been demonstrated that our method can successfully learn an image-to-image translation model from noisy medical imaging measurements for establishing realistic SOMs that possess the ability to control features of the synthesized objects. 
	
\section{Method}
	
\subsection{CycleGAN}
\label{sec:title}
CycleGAN learns to translate image data between the source domain $X$ and the target domain $Y$. It consists of two generator networks, $G_x$ and $G_y$, where $G_x$ maps image data from the domain $X$ to the domain $Y$ and $G_y$ maps image data from the domain $Y$ to the domain $X$. CycleGAN also employs two discriminator networks, $D_x$ and $D_y$, acting on image data in the domain $X$ and $Y$, respectively. The training aims to minimize the loss function $\mathcal{L}(G_x, G_y, D_x, D_y)$:
\begin{equation}
\mathcal{L}(G_x,G_y,D_x,D_y)=\mathcal{L}_{\mathrm{GAN}}(G_x,D_y,X,Y)+\mathcal{L}_{\mathrm{GAN}}(G_y,D_x,Y,X)+\lambda\mathcal{L}_{\mathrm{cyc}}(G_x,G_y),
\end{equation}
where $\mathcal{L}_{\mathrm{GAN}}(G_x, D_y, X, Y)$ and $\mathcal{L}_{\mathrm{GAN}}(G_y, D_x, Y, X)$ are the adversarial losses, and $\mathcal{L}_{\mathrm{cyc}}(G_x, G_y)$ is the cycle consistency loss. The cycle consistency loss is to ensure that the mappings $G_x$ and $G_y$ produce the translated images that possess consistent structures to those of the input images. Here, $\lambda$ controls the relative importance of the cycle consistent loss to the adversarial losses. Further information on CycleGAN can be found in the origin paper \cite{CycleGAN}. However, CycleGAN cannot be trained directly on noisy measurement data for establishing SOMs because SOMs should only characterize object variability.

\subsection{AmbientCycleGAN}

\begin{figure}[h!]
\centering
\includegraphics[width=\linewidth]{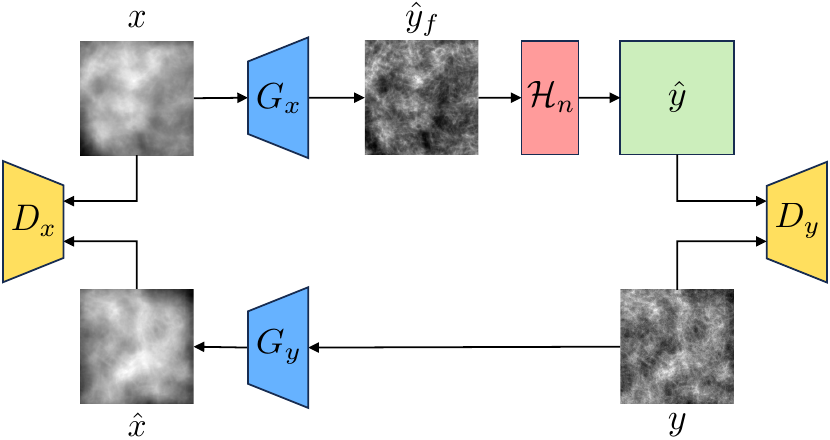}
\caption{Illustration of the proposed AmbientCycleGAN architecture for establishing realistic SOMs based on mathematical phantoms and noisy imaging measurements.}
\label{Am}
\end{figure}

Bora \emph{et al.} introduced AmbientGAN \cite{Ambientgan} for learning object distribution from noisy data. More recently, Zhou \emph{et al.} explored its application in creating SOMs from measured imaging data \cite{zhouAmbJMI}. However, it still remains unclear how AmbientGAN can be employed for establishing SOMs that can produce objects in an interpretable way. In this work, we propose a novel AmbientCycleGAN that can translate mathematical SOMs to realistic SOMs by use of noisy measurement data. Together, the utilized mathematical SOM and the AmbientCycleGAN represent an SOM that can produce realistic objects and interpretably control object content.

As shown in Fig. \ref{Am}, similar to CycleGAN, AmbientCycleGAN consists of two generators $G_x$ and $G_y$, where generator $G_y$ maps noisy measurement data $y$ to the synthesized objects $\hat{x}$ corresponding to the utilized mathematical SOM. Meanwhile, $G_x$ maps objects $x$ produced by the mathematical SOM to the synthesized objects $\hat{y}_f$ corresponding to the noisy measurement data. Subsequently, the measurement operator $\mathcal{H}_n$ is applied to $\hat{y}_f$ to compute the corresponding simulated imaging measurement data $\hat{y}$. Regarding the two discriminators, $D_x$ compares $\hat{x}$ with the real samples $x$, and $D_y$ discriminates between the real measurements $y$ and the generated measurements $\hat{y}$. The corresponding adversarial and cycle consistency loss functions are described as:
\begin{align}
&\mathcal{L}_{\mathrm{GAN}}(G_x,D_{y},X,Y) =\mathbb{E}_{y\in \mathbf{Y}}[\log D_{y}(y)] +\operatorname{E}_{x \in \mathbf{X}}[\log(1-D_{y}(\mathcal{H}_n(G_x(x)))] \\
&\mathcal{L}_{\mathrm{GAN}}(G_y,D_{x},Y,X) =\mathbb{E}_{x\in \mathbf{X}}[\log D_{x}(x)] +\operatorname{E}_{y \in \mathbf{Y}}[\log(1-D_{x}(G_y(y))] \\
 &\mathcal{L}_{\mathrm{cyc}}(G_x,G_y)=\mathbb{E}_{x\in \mathbf{X}}[\|G_y(\mathcal{H}_n(G_x(x)))-x\|_{2}]+\mathbb{E}_{y \in \mathbf{Y}}[\|\mathcal{H}_n(G_x(G_y(y)))-y\|_{2}]
\end{align}

\section{Numerical studies}
\label{sec:sections}
	

Two cases were considered in our numerical studies. The first case involved two modified CLB object models, Opex-CLB and Simpiso-CLB \cite{castella2008mammographic}, where Opex-CLB represented the pre-existing mathematical SOM and Simpiso-CLB was considered as the ground-truth SOM from which we can obtain noisy measurement data. To form training dataset, 20,000 images of size $256 \times 256$ were produced by Opex-CLB and Simpiso-CLB, and Gaussian noise (mean 0, standard deviation (std) 0.04) was added to the Simpiso-CLB-produced objects to simulate the noisy measurement data. 

In a more realistic scenario, real mammograms were employed. Specifically, Opex-CLB still served as the mathematical SOM to be input to the translation model, and two mammography databases, DDSM \cite{heath1998current} and CBIS-DDSM \cite{lee2017curated}, were used to form a noisy measurement dataset corresponding to the ground-truth SOM. Images from DDSM and CBIS-DDSM were both resized to 256 $\times$ 256. A total number of 13,190 real mammograms were subsequently collected and this new dataset will be referred to as ``DDSM/CBIS-DDSM'' in the rest of this paper. Similarly, Gaussian noise (mean 0, std 0.04) was added to this dataset to emulate realistic noisy measurement data.

The AmbientCycleGANs were trained by use of PyTorch on an NVIDIA RTX 3090 Ti GPU. It was implemented based on the CycleGAN framework: \href{https://github.com/junyanz/pytorch-CycleGAN-and-pix2pix/}{https://github.com/junyanz/pytorch-CycleGAN-and-pix2pix}. The Adam optimizer with a learning rate of 0.0002 was deployed in the experiments. Moreover, a Res-Net architecture \cite{CTCylceGAN} was employed in the generator architecture, and the batch size was set to 10.  
\section{Results}

\begin{figure}[!ht]
		\centering
		\includegraphics[width=\linewidth]{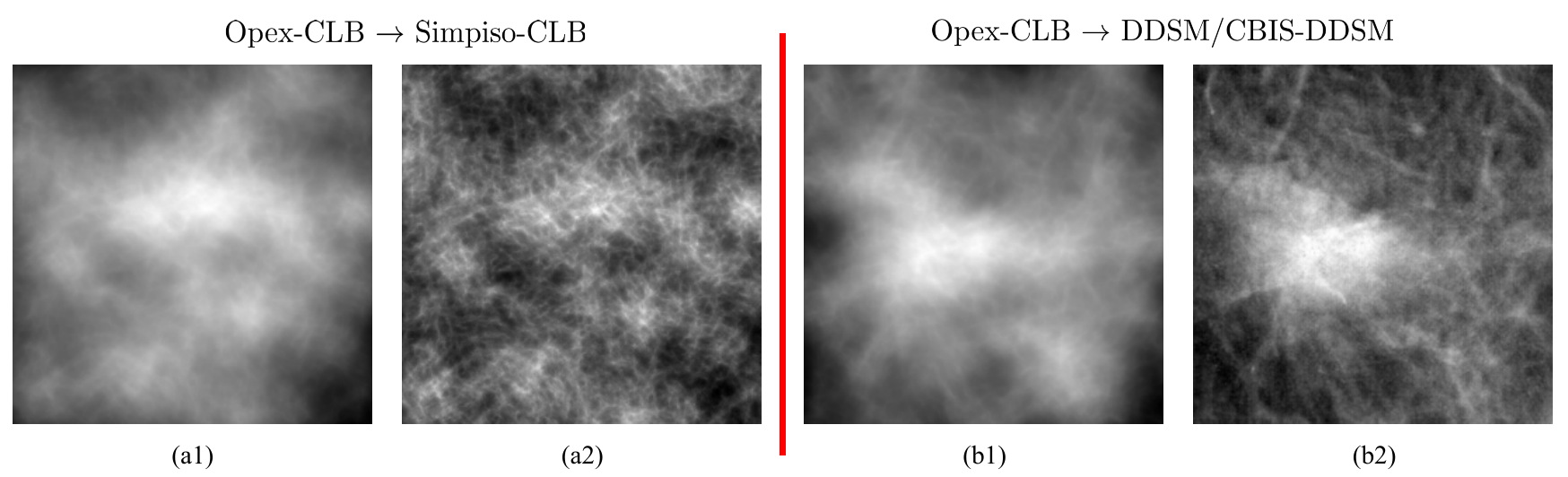}
		\caption{(a1) and (b1) are Opex-CLB objects that input to the AmbientCycleGAN, (a2) and (b2) are the AmbientCyleGAN-generated objects corresponding to SOMs represented by the CLB-Simpiso and DDSM/CBIS-DDSM.}
		\label{dataset}
\end{figure}

\begin{figure}[ht!]
		\centering
		\includegraphics[width=\linewidth]{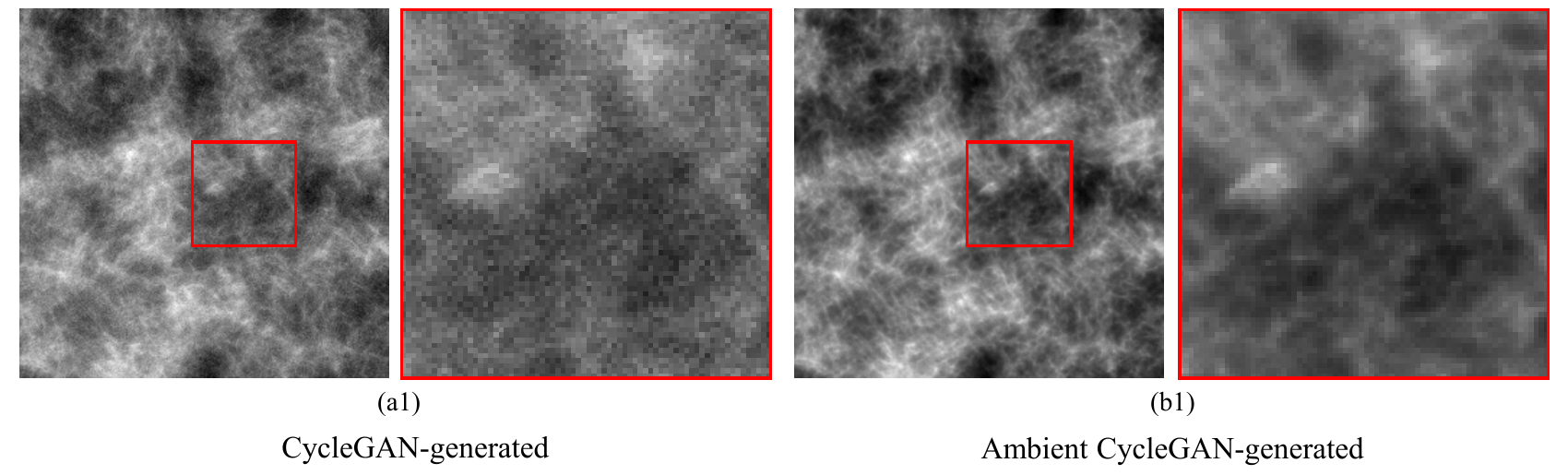}\\
		\caption{Images generated by the traditional CycleGAN (left) and images generated by the proposed AmbientCycleGAN (right) in the Opex-CLB and Simpiso-CLB case.}
		\label{detail}
	\end{figure}

\begin{figure}[ht!]
		\centering
		\includegraphics[width=\linewidth]{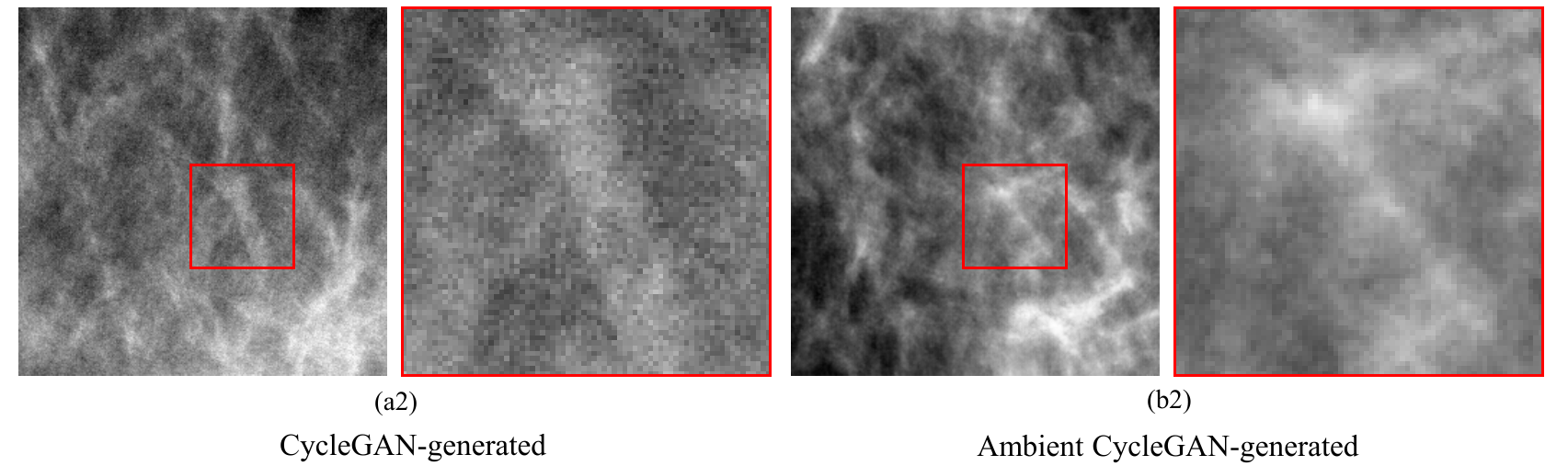}\\
		\caption{Images generated by the traditional CycleGAN (left) and images generated by the proposed AmbientCycleGAN (right) in the Opex-CLB and DDSM/CBIS-DDSM case.}
		\label{detail2}
	\end{figure}

Examples of AmbientCycleGAN-produced objects are shown in 
 Fig. \ref{dataset}, from which the AmbientCycleGAN-produced images had an appearance that is consistent with the real measurement data and the overall structure is consistent with the input mathematical phantom. Additionally, images produced by the AmbientCycleGAN are compared with those produced by the standard CycleGAN in Fig. \ref{detail} and Fig. \ref{detail2} for the two considered cases. It is observed that the CycleGAN-produced images were strongly affected by measurement noise; while the AmbientCycleGAN-produced images were clean. 

	The Frechlet Inception Distance (FID) score, a widely used metric for assessing GANs, was employed to evaluate the AmbientCycleGAN-generated images based on 6000 images. Our results, presented in Table \ref{fid}, suggest that AmbientCycleGAN achieved significantly lower FID scores compared to CycleGAN across both datasets, indicating superior image quality and performance of AmbientCycleGAN.
	\begin{table}[htb]
		\centering
		\begin{tabular}{ccc}
			\toprule
			\multirow{1}{*}{Dataset}
		&	\multirow{1}{*}{Method}  
			& \multicolumn{1}{c}{FID ($\downarrow$)} \\ \midrule
		\multirow{2}*{Opex-CLB \& Simpiso-CLB} &	CycleGAN                 &           127.69                     \\
		&AmbientCycleGAN    &                13.91                  \\ 
			\midrule
		\multirow{2}*{Opex-CLB \& DDSM/CBIS-DDSM} &	CycleGAN                 &           82.23                     \\
		&AmbientCycleGAN    &                17.89                  \\
			\bottomrule
		\end{tabular}

		\caption{FID scores of the AmbientCycleGAN- and CycleGAN-generated images on two datasets.}
	\label{fid}
	\end{table}
		
\begin{figure}[htb!]
		\centering
		
		\includegraphics[width=\linewidth]{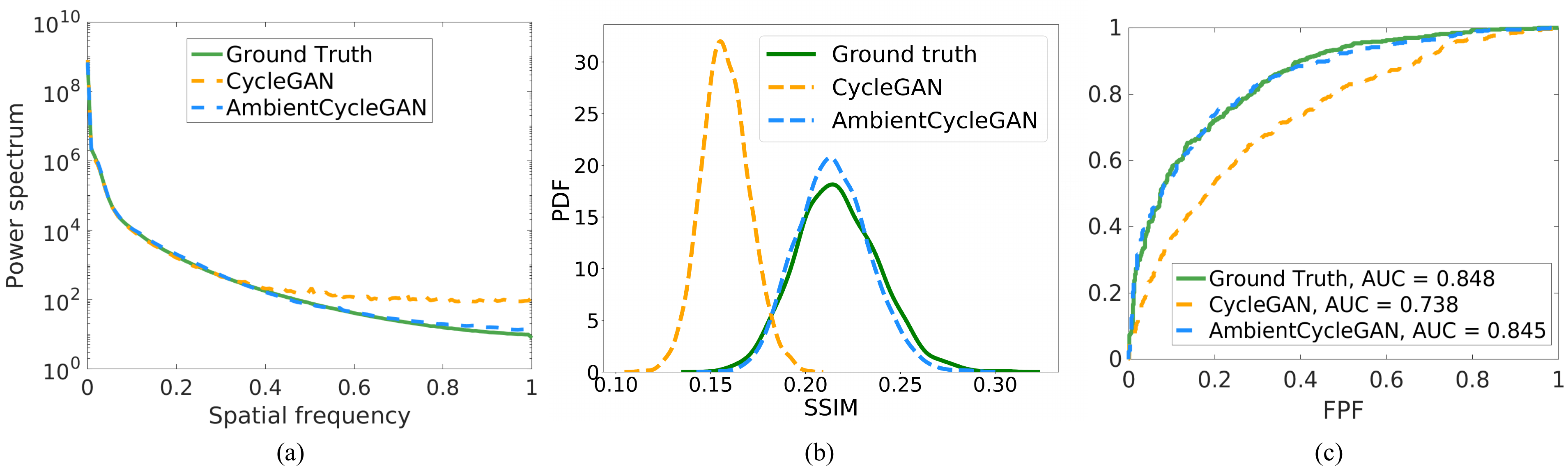}\\
		\caption{ (a) Radially averaged power spectrum, (b) PDFs of SSIMs, (c) ROC curves corresponding to the Opex-CLB \& Simpiso-CLB case.}
		\label{d2img2}
	\end{figure}

\begin{figure}[htb!]
		\centering
		
		\includegraphics[width=\linewidth]{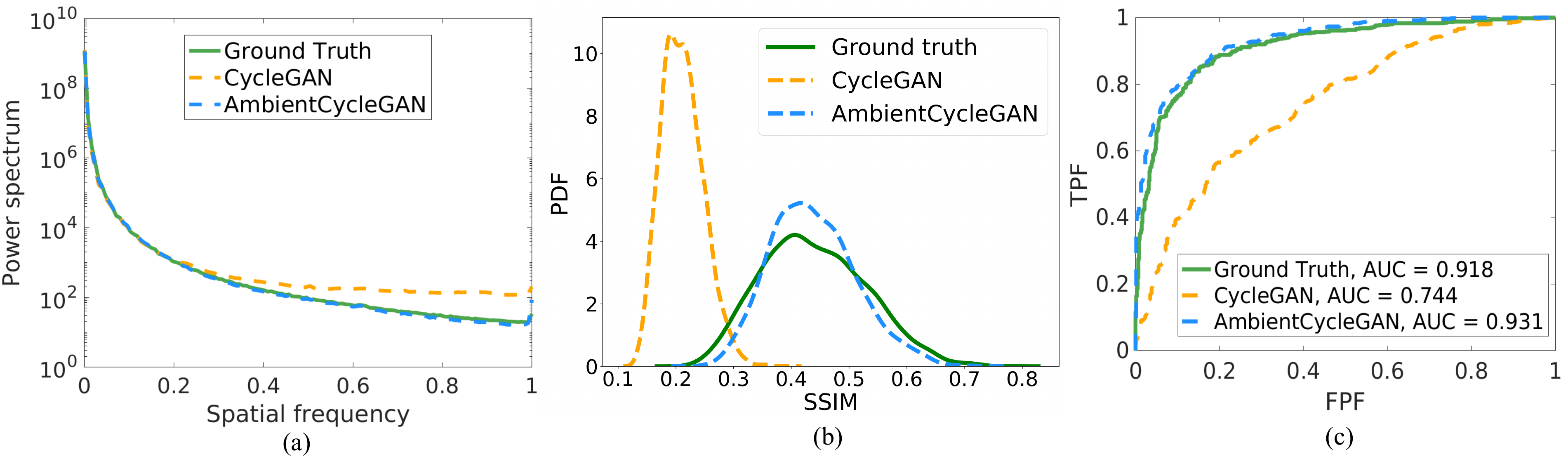}\\
		\caption{(a) Radially averaged power spectrum, (b) PDFs of SSIMs, (c) ROC curves corresponding to the Opex-CLB \& DDSM/CBIS-DDSM case.}
		\label{img_ddsm}
	\end{figure}

To further assess the performance of the AmbientCycleGAN, the radially averaged power spectrum of 5000 synthesized images produced by CycleGAN and AmbientCycleGAN were compared to those of the ground-truth images. As shown in Fig. \ref{d2img2} (a) and Fig. \ref{img_ddsm} (a), the AmbientCycleGAN-generated images produced a power spectrum that was similar to that of the ground truth; while CycleGAN-generated images produced a power spectrum having noticeable differences to the ground truth.

Additionally, the Structure Similarity Index Measure (SSIM) values for random pairs of ground truth objects, random pairs of AmbientCycleGAN-generated and ground truth objects, and random pairs of CycleGAN-generated and ground truth objects were calculated. Probability Density Functions (PDFs) of these SSIMs were compared in Fig. \ref{d2img2} (b) and Fig. \ref{img_ddsm} (b). The PDFs corresponding to the AmbientCycleGAN and the ground truth PDFs were closely matched; while the PDFs corresponding to the CycleGAN had significant divergence from the ground-truth PDFs, as expected.

Moreover, a task-based image quality study was conducted to evaluate AmbientCycleGAN's ability to establish SOMs for signal detection tasks. This involved a signal-known-exactly binary signal detection task using the ground truth SOM-, CycleGAN- and AmbientCycleGAN-produced background images. We first cropped the $64 \times 64$ patches from the central region of each background image. Subsequently, a Gaussian signal with a mean of 0.3 and a standard deviation of 0.7 was superimposed onto these patches to simulate signal-present images. The Hotelling observer (HO) was computed to evaluate the signal detection performance. A set of 4,500 objects was used for computing the HO template and  a set of 1,000 images was used for evaluating the detection performance. The receiver operating characteristic (ROC) curves shown in Fig. \ref{d2img2} (c) and Fig. \ref{img_ddsm} (c) suggest that the HO performance corresponding to the AmbientCycleGAN-represented SOMs closely resembled that corresponding to the ground-truth images; while the ROC curves evaluated on the standard CycleGAN-produced images had a significant discrepancy to the ground-truth ROC curve.

\begin{figure}[ht!]
		\centering
		\includegraphics[width=\linewidth]{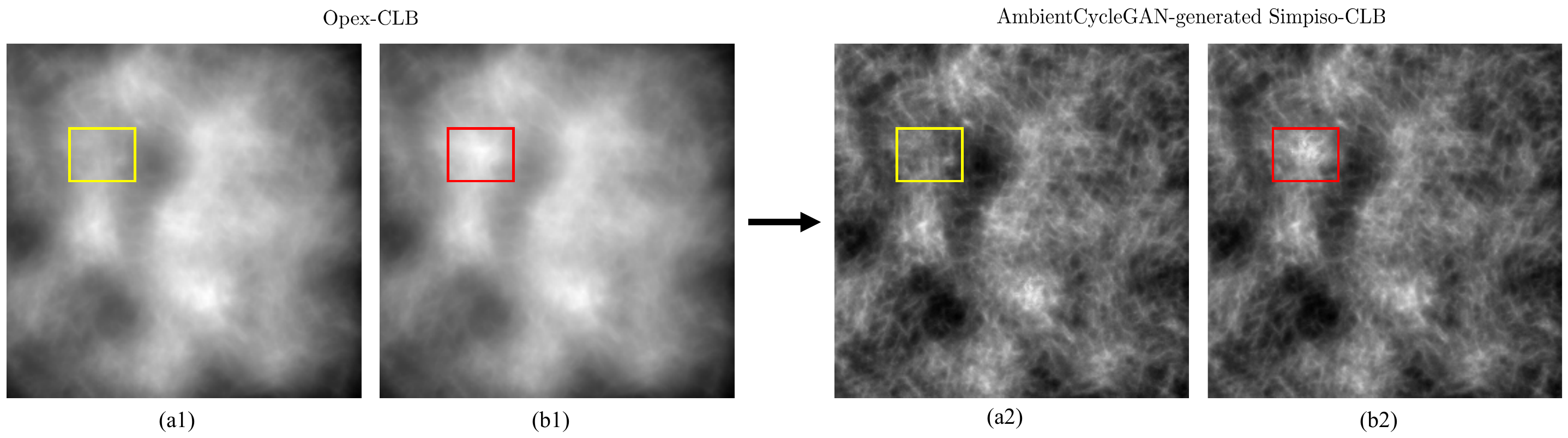}\\
		\caption{(a1) and (b1) are the Opex-CLB-generated objects that differ in the number of clusters in the yellow and red boxes-circled region, (a2) and (b2) are the corresponding AmbientCycleGAN-translated images in Simpiso-CLB domain.}
		\label{cluster}
	\end{figure}

\begin{figure}[h!]
		\centering
		\includegraphics[width=\linewidth]{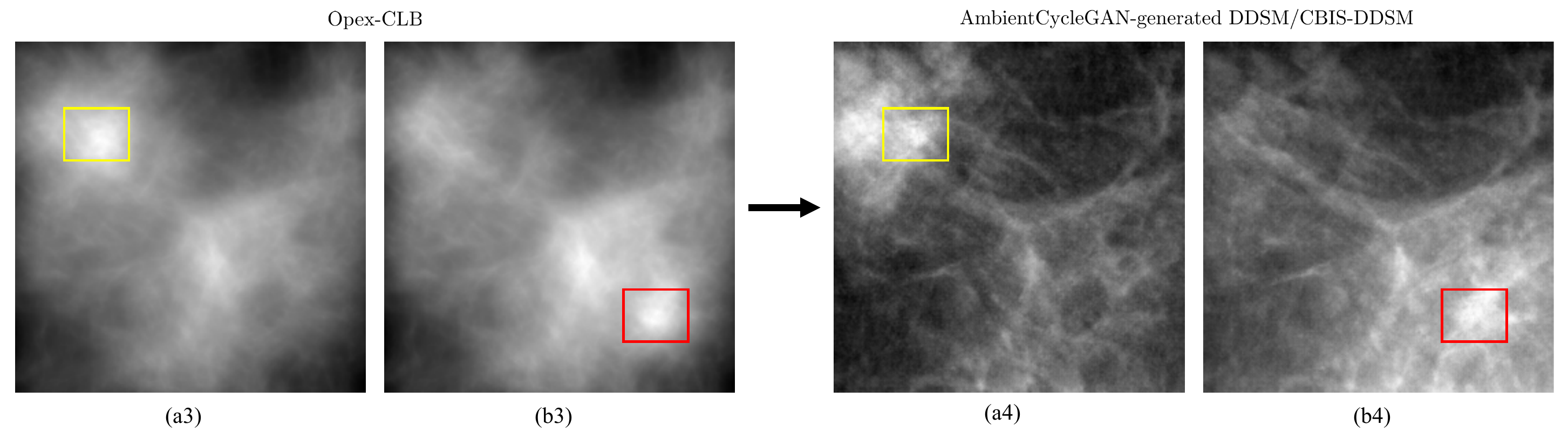}\\
		\caption{(a3) and (b3) are the Opex-CLB-generated objects that differ in the number of clusters in the yellow and red boxes-circled region, (a4) and (b4) are the corresponding AmbientCycleGAN-translated mammograms.}
		\label{cluster2}
	\end{figure}

We also demonstrate that the proposed AmbientCycleGAN can control object features in an interpretable way. As shown in Fig. \ref{cluster}, additional cluster of lumps were added to the yellow box-circled region in the input Opex-CLB image (a1) to form a new Opex-CLB image (b1). This action led to the corresponding addition of the cluster of lumps in the same region in the AmbientCycleGAN-generated Simpiso objects, as shown in (a2) and (b2). In another case, we moved the top-left cluster of lumps in Fig. \ref{cluster2} (a3) to a bottom-right location (b3), and observed that
in the AmbientCycleGAN-produced mammograms, the dense breast tissues also moved from a top-left region to a bottom-right region.

\section{Conclusions}
In this work, a deep learning approach that employs AmbientCycleGAN was proposed for establishing interpretable stochastic object models (SOMs) based on pre-existing mathematical phantoms and medical imaging measurements. Preliminary studies that considered the clustered lumpy background (CLB) models and clinic mammograms have demonstrated that the proposed AmbientCycleGAN can be effectively trained on noisy measurement data to translate simple SOMs to realistic SOMs. It was also demonstrated that the AmbientCycleGAN-established SOMs possess the ability to manipulate image features in an interpretable way. Such SOMs with interpretable controls represent an useful tool that may enable a more comprehensive analysis of medical imaging systems that involves a variety of controlled object variability.
	
	\bibliography{report} 
	\bibliographystyle{spiebib} 
	
\end{document}